\documentclass[aps,prd,twocolumn,showpacs,superscriptaddress,groupedaddress,nofootinbib]{revtex4}  
\usepackage[a4paper]{geometry}
\usepackage{graphicx}  
\usepackage{dcolumn}   
\usepackage{bm}        
\usepackage{amssymb}   
\usepackage{amsmath}
\usepackage{array}
\usepackage{color}
\usepackage{rotating}
\usepackage{footnote}
\newcommand{\PreserveBackslash}[1]{\let\temp=\\#1\let\\=\temp}
\newcolumntype{C}[1]{>{\PreserveBackslash\centering}p{#1}}
\newcolumntype{R}[1]{>{\PreserveBackslash\raggedleft}p{#1}}
\newcolumntype{L}[1]{>{\PreserveBackslash\raggedright}p{#1}}

\usepackage{multirow}

\newcommand{\llb}{\Lambda\bar{\Lambda}}
\newcommand{\ssb}{\Sigma^0\bar{\Sigma}^0}
\newcommand{\Nj}{1310.6\times10^6} 
\newcommand{\Np}{447.9\times10^6} 
\newcommand{\Njpsi}{(1310.6\pm7.0)\times10^6} 
\newcommand{\Npsip}{(447.9\pm2.9)\times10^6} 

\newcommand{\ai}{0.469\pm0.026\pm0.008}  
\newcommand{\aii}{-0.449\pm0.020\pm0.008}  
\newcommand{\aiii}{0.82\pm0.08\pm0.02}  
\newcommand{\aiv}{0.71\pm0.11\pm0.04}  
\newcommand{\bi}{19.43\pm0.03\pm0.33}  
\newcommand{\bii}{11.64\pm0.04\pm0.23}  
\newcommand{\biii}{3.97\pm0.02\pm0.12}  
\newcommand{\biv}{2.44\pm0.03\pm0.11}  
\newcommand{\aip}{0.469\pm0.026}
\newcommand{\aiip}{-0.449\pm0.020}
\newcommand{\aiiip}{0.824\pm0.074}
\newcommand{\aivp}{0.71\pm0.11}
\newcommand{\bip}{19.43\pm0.03}
\newcommand{\biip}{11.64\pm0.04}
\newcommand{\biiip}{3.97\pm0.02}
\newcommand{\bivp}{2.44\pm0.03}
\newcommand{\qi}{20.43 \pm0.11 \pm0.58}     
\newcommand{\qii}{20.96 \pm0.27  \pm0.92}    

\begin{document}

\title{\boldmath Study of $J/\psi$ and $\psi(3686)$ decay to $\llb$ and $\ssb$ final states}

\author{
M.~Ablikim$^{1}$, M.~N.~Achasov$^{9,e}$, S. ~Ahmed$^{14}$, M.~Albrecht$^{4}$, A.~Amoroso$^{53A,53C}$, F.~F.~An$^{1}$, Q.~An$^{50,a}$, J.~Z.~Bai$^{1}$, Y.~Bai$^{39}$, O.~Bakina$^{24}$
% ...
\\
\vspace{0.2cm}
(BESIII Collaboration)\\
\vspace{0.2cm} {\it
$^{1}$ Institute of High Energy Physics, Beijing 100049, People's Republic of China\\
% ...
$^{4}$ Bochum Ruhr-University, D-44780 Bochum, Germany\\
% ...
$^{9}$ G.I. Budker Institute of Nuclear Physics SB RAS (BINP), Novosibirsk 630090, Russia\\
% ...
$^{14}$ Helmholtz Institute Mainz, Johann-Joachim-Becher-Weg 45, D-55099 Mainz, Germany\\
% ...
$^{24}$ Joint Institute for Nuclear Research, 141980 Dubna, Moscow region, Russia\\
% ...
$^{39}$ Southeast University, Nanjing 211100, People's Republic of China\\
% ...
$^{50}$ University of Science and Technology of China, Hefei 230026, People's Republic of China\\
% ...
$^{53}$ (A)University of Turin, I-10125, Turin, Italy; (B)University of Eastern Piedmont, I-15121, Alessandria, Italy; (C)INFN, I-10125, Turin, Italy\\
% ...
$^{a}$ Also at State Key Laboratory of Particle Detection and Electronics, Beijing 100049, Hefei 230026, People's Republic of China\\
% ...
$^{e}$ Also at the Novosibirsk State University, Novosibirsk, 630090, Russia\\
% ...
}
}

\date{\today}

\begin{abstract}
  Using $\Nj$ $J/\psi$ and $\Np$ $\psi(3686)$ events collected with
  the BESIII detector at the BEPCII $e^{+}e^{-}$ collider, the
  branching fractions and the angular distributions of $J/\psi$ and
  $\psi(3686)$ decays to $\llb$ and $\ssb$ final states are measured.
  The branching fractions are determined, with much improved precision,
  to be $\bi$, $\bii$, $\biii$ and $\biv$ for $J/\psi\to\llb$, $J/\psi\to\ssb$, 
  $\psi(3686)\to\llb$ and $\psi(3686)\to\ssb$, respectively. The polar
  angular distributions of $\psi(3686)$ decays are measured for the
  first time, while those of $J/\psi$ decays are measured with much
  improved precision.  In addition, the ratios of branching fractions
  $\frac{\mathcal{B}(\psi(3686)\to\llb)}{\mathcal{B}(J/\psi\to\llb)}$
  and
  $\frac{\mathcal{B}(\psi(3686)\to\ssb)}{\mathcal{B}(J/\psi\to\ssb)}$
  are determined to test the ``12\% rule''.
\end{abstract}


\maketitle

\section{\bf INTRODUCTION} \label{Sec:intro}

Two-body baryonic decays of $\psi$ mesons ($\psi$ denotes both the
$J/\psi$ and $\psi(3686)$ charmonium states throughout the text), take
place through annihilation of the constituent $c\bar{c}$ quark
pair into either a virtual photon or three gluons, and they provide a
good laboratory for testing Quantum Chromodynamics (QCD) in the
perturbative energy regime and studying the properties of
baryons~\cite{Ref:intro}. Perturbative QCD (pQCD) predicts that the
ratio of branching fractions between the $J/\psi$ and $\psi(3686)$
decaying into a given hadronic final states follows the ``12\%
rule"~\cite{Ref:12rule}
\begin{equation}
\label{equ:.12rule}
Q = \frac{\mathcal{B}_{\psi(3686)\to h}}{\mathcal{B}_{J/\psi\to h}} = \frac{\mathcal{B}_{\psi(3686)\to l^+l^-}}{\mathcal{B}_{J/\psi\to l^+l^-}} \approx (12.4 \pm 0.4)\%.
\end{equation}
The violation of this rule was first observed in the decay of $\psi$
into the final state $\rho\pi$, which is well known as the ``$\rho\pi$
puzzle"~\cite{Ref:rhopi_puzzle}, and the rule has been subsequently
further tested in a wide variety of experimental
measurements.  Reviews of the theoretical and
experimental results~\cite{Ref:review} conclude that the current
theoretical understanding, especially for the $\psi$ decays into
baryon-antibaryon pair final states, is not mature. The branching
fractions of $\psi$ decays into $B\bar{B}$ ($B\bar{B}$ refers to both
$\llb$ and $\ssb$ throughout the text) final states from different
experiments~\cite{Ref:markii_1984, Ref:dm2_1987,Ref:bes_1998,
  Ref:bes_2001, Ref:cleo_2005, Ref:bes_2006, Ref:bes_2007,
  Ref:babar_2007, Ref:cleo_2014, Ref:bes_2016} and the Particle Data
Group (PDG)~\cite{Ref:pdg} averages are summarized in
Table~\ref{Tab:br_list}.  Obvious differences between the different
experiments are observed, and the uncertainties are relatively
large. Hence, higher precision measurements of the $\psi$ decays into
$B\bar{B}$ pairs are desirable to help in understanding the dynamics
of $\psi$ decay.

\begin{table*}[!htb]
\begin{center}
\caption{\label{Tab:br_list}\small Experimental measurements and PDG averages for the branching fractions of the decay $\psi\to B\bar{B}$ $(\times10^{-4})$.}
\begin{tabular}{ccccc}\hline\hline
                                                                & $J/\psi\to\llb$ & $\psi(3686)\to\llb$ & $J/\psi\to\ssb$ & $\psi(3686)\to\ssb$ \\\hline
MARKII Collab.~\cite{Ref:markii_1984}  &     $15.8\pm0.8\pm1.9$ & ... &  $15.8\pm1.6\pm2.5$ & ... \\
DM2 Collab.~\cite{Ref:dm2_1987}  &     $13.8\pm0.5\pm2.0$ & ... & $10.6\pm0.4\pm2.3$ & ... \\
BES Collab.~\cite{Ref:bes_1998, Ref:bes_2001} &     $10.8\pm0.6\pm2.4$ & $1.8\pm0.2\pm0.3$ & ... & $1.2\pm0.4\pm0.4$ \\
CLEO Collab.~\cite{Ref:cleo_2005} & ... &     $3.3\pm0.3\pm0.3$ & ... & $2.6\pm0.4\pm0.4$ \\
BESII Collab.~\cite{Ref:bes_2006, Ref:bes_2007} &     $20.3\pm0.3\pm1.5$ & $3.4\pm0.2\pm0.4$ & $13.3\pm0.4\pm1.1$ & $2.4\pm0.4\pm0.4$ \\
BaBar Collab.~\cite{Ref:babar_2007} &     $19.3\pm2.1\pm0.5$ & $6.4\pm1.8\pm0.1$ & $11.5\pm2.4\pm0.3$ & ... \\
S.~Dobbs {\it et al.}~\cite{Ref:cleo_2014} &     ... & $3.8\pm0.1\pm0.3$ & ... & $2.3\pm0.2\pm0.2$ \\\hline
PDG~\cite{Ref:pdg} &     $16.1\pm1.5$ & ~$3.6\pm0.2$ & $12.9\pm0.9$ & $2.3\pm0.2$ \\
\hline\hline
\end{tabular}
\end{center}
\end{table*}

The angular distribution of the decays $e^+e^-\to\psi\to B\bar{B}$ can
be expressed in form~\cite{Ref:intro}
\begin{equation}
\label{equ:angdis}
\frac{dN}{d\cos\theta}\propto1+\alpha \cos^2\theta,
\end{equation}
where $\theta$ is the angle between the outgoing baryon and the beam
direction in the $e^+e^-$ center-of-mass (c.m.) system, and $\alpha$
is a constant, which is related to the decay properties.  The equation
is derived from the general helicity formalism~\cite{Ref:intro},
taking into account the gluon spin, the quark distribution amplitudes
in $e^+e^-\to\psi\to B\bar{B}$, and hadron helicity conservation. The
$\alpha$ values in the decays $J/\psi\to B\bar{B}$ have been
calculated with pQCD to first-order~\cite{Ref:claudson}. It is
believed that the masses of the baryon and quark must be taken into
consideration in the $\alpha$ calculation since a large violation of
helicity conservation is observed in $\psi$ decays~\cite{Ref:claudson,
  Ref:carimalo}. Table~\ref{Tab:alpha_list} summarizes the theoretical
predictions and experimental $\alpha$ values for the
decays $J/\psi\to B\bar{B}$. To date, the experimental $\alpha$ values for
the decays $J/\psi\to B\bar{B}$ have poor
precision~\cite{Ref:markii_1984,Ref:dm2_1987,Ref:bes_2006}, and the
alpha values in the decay $\psi(3686)\to B\bar{B}$ have not
yet been measured. It is worth noting that there is an indication that
the $\alpha$ value in the decay $J/\psi\to\ssb$ is negative in
Ref.~\cite{Ref:bes_2006}.

\begin{table}[!htb]
\caption{\label{Tab:alpha_list}\small Theoretical predictions  and experimental measurements of $\alpha$ for $J/\psi\to B\bar{B}$.}
\begin{center}
\begin{tabular}{cccc}\hline\hline
&$\alpha_{J/\psi\to\llb}$ & $\alpha_{J/\psi\to\ssb}$\\
\hline
\multirow{2}{2cm}{Theory}&0.32 &  0.31~\cite{Ref:claudson}\\
&0.51 &  0.43~\cite{Ref:carimalo}\\
\hline
\multirow{3}{2cm}{Experiment}&$0.72\pm0.36$ & $0.70\pm1.10$~\cite{Ref:markii_1984}\\
&$0.62\pm0.22$ & $0.22\pm0.31$~\cite{Ref:dm2_1987}\\
&$0.65\pm0.14$ & $-0.22\pm0.19$~\cite{Ref:bes_2006}\\
\hline\hline
\end{tabular}
\end{center}
\end{table}

\vspace{-1mm}
In this paper, we report precise measurements of the branching
fractions and $\alpha$ values for the decays $\psi\to B\bar{B}$,
based on the data samples of $\Njpsi$ $J/\psi$~\cite{Ref:ntot_jpsi}
and $\Npsip$ $\psi(3686)$~\cite{Ref:ntot_psip} events collected with
the BESIII detector at the BEPCII collider.

\section{BESIII DETECTOR AND DATA SET}\label{Sec:detector}
The BESIII detector~\cite{Ref:bes3_detector} at the double-ring
Beijing Electron-Positron Collider~(BEPCII)~\cite{Ref:bepcii} is
designed for studies of physics in the $\tau$-charm energy
region~\cite{Ref:yellowbook}.
The peak luminosity of BEPCII is $10^{33}$~cm$^{-2}$~s$^{-1}$ at a
beam current of 0.93~A. The BESIII detector has a geometrical
acceptance of 93\% of $4\pi$ solid angle and consists of the following
main components:
(1)~A small-celled, helium based (40\% CO$_2$ and 60\% C$_3$H$_8$)
main drift chamber~(MDC) with 43 layers, which has an average
single-wire resolution of 135~$\mu$m, a momentum resolution for
1~GeV/c charged particles in a 1~T magnetic field of 0.5\%
\renewcommand{\thefootnote}{\fnsymbol{footnote}}
\setcounter{footnote}{1}
\footnote{For the $J/\psi$ data sample collected in 2012, the magnetic field was 0.9~T.}, 
and a specific energy loss ($dE/dx$) resolution of better than 6\%.
(2)~An electromagnetic calorimeter~(EMC), which consists of 6240 CsI~(Tl)
crystals arranged in a cylindrical shape (barrel) plus two
end-caps. For 1.0~GeV photons, the energy resolution is 2.5\% (5\%) in
the barrel (end-caps), and the position resolution is 6~mm (9~mm) for
the barrel (end-caps).
(3)~A time-of-flight (TOF) system, which is used for particle
identification (PID). It is composed of a barrel made of two layers,
each consisting of 88 pieces of 5~cm thick and 2.4~m long plastic
scintillators, as well as two end-caps each with 96 fan-shaped 5~cm
thick plastic scintillators. The time resolution is 80~ps (110~ps) in
the barrel (end-caps), providing a $K/\pi$ separation of more than
2$\sigma$ for momenta up to 1.0~GeV/c.
(4)~A muon chamber system, which is made of resistive plate chambers
(RPCs) arranged in 9 layers (8 layers) in the barrel (end-caps) with
$\sim$ 2 cm position resolution. It is incorporated into the return
iron yoke of the superconducting magnet.

The optimization of the event selection and the estimations of the
signal detection efficiency and background are determined using Monte
Carlo (MC) simulations. The GEANT4-based~\cite{Ref:geant4} simulation
software BOOST~\cite{Ref:boost}, which includes the geometric and
material description of the BESIII detector, the detector response and
digitization models, as well as the tracking of the detector running
conditions and performance, is used to generate MC samples. The
analysis is performed in the framework of the BESIII offline software
system (BOSS)~\cite{Ref:boss} which takes care of the detector
calibration, event reconstruction and data storage.

Generic inclusive MC samples, which include $1,225\times10^6$
$J/\psi$ and $460\times10^6$ $\psi(3686)$ events, are used to study
the potential backgrounds. The $\psi$ are produced via
$e^{+}e^{-}\to\psi$ processes by the generator KKMC~\cite{Ref:kkmc},
which includes the beam energy spread according to the measurement of
BEPCII and the effect of initial state radiation (ISR). The known
decay modes are generated with BesEvtGen~\cite{Ref:evtgen} according
to world average branching fraction values~\cite{Ref:pdg}; the
remaining unknown decay modes are simulated using the LundCharm
model~\cite{Ref:lundcharm}. To determine the detection efficiencies,
large $\psi\to$ $B\bar{B}$ signal MC samples are generated for each
process, where the angular distributions of the baryons use $\alpha$
values obtained in this analysis. The $\Lambda$ and $\Sigma^{0}$
particles are simulated in the $\Lambda\to p \pi^{-}$ and
$\Sigma^{0}\to\gamma\Lambda$ decay modes.

%
%
\section{\bf EVENT SELECTION } \label{Sec:evtslt}
In this analysis, the four decay modes $\psi\to B\bar{B}$ are studied
by fully reconstructing both $B$ and $\bar{B}$, where the
$\Lambda(\bar{\Lambda})$ and $\Sigma^{0}(\bar{\Sigma^0})$ candidates
are reconstructed with the $ p \pi^{-}(\bar{p}\pi^+)$ and
$\gamma\Lambda(\gamma \bar{\Lambda})$ decay modes, respectively. Therefore,
the decays $\psi\to\llb$ and $\psi\to\ssb$ have the final states
$p\bar{p}\pi^+\pi^-$ and $p\bar{p}\pi^+\pi^-\gamma\gamma$,
respectively.

Events with at least four charged tracks with total charge zero are
selected. Each charged track is required to have
$|\cos \theta|<0.93$, where $\theta$ is the polar angle of the
track.
Photons are reconstructed from isolated showers in the EMC which are
at least 30 degrees away from the anti-proton and 10 degrees from
other charged tracks. The energy deposited in the nearby TOF counters
is included to improve the photon reconstruction efficiency and energy
resolution. Photon candidates are required to be within the barrel
region ($|\cos \theta|<0.8$) of the EMC with deposited energy of
at least 25~MeV, or within the end cap regions
($0.86<|\cos \theta|<0.92$) with at least 50~MeV, where $\theta$
is the polar angle of the photon.  In order to suppress electronic
noise and energy deposits unrelated to the event, the timing
information $t$ from the EMC for the photon candidate must be in
coincidence with the collision event ($0\le t\le700$~ns).  At least
two photons are required in the analysis of $\psi\to\ssb$ decays.

MC studies indicate that the proton and pion from $\Lambda$ decay are
well separated kinematically since the proton carries most of the
energy. A charged track with momentum $p>0.5$~GeV/c is assumed to be a
proton, while that with $p<0.5$~GeV/c is assumed to be a pion. The
$\Lambda$ ($\bar{\Lambda}$) candidate is reconstructed with any
$p\pi^{-}$ ($\bar{p}\pi^{+}$) combination satisfying a secondary
vertex fit~\cite{Ref:secVFit} and having a decay length larger 
than 0.2 cm to suppress the non-$\Lambda$ (non-$\bar{\Lambda}$) decays.
The decay length is the distance between its primary vertex and decay point
to $p\pi^{-}$ ($\bar{p}\pi^{+}$), where the primary vertex is approximated by 
the interaction point averaged over many events. If more than
one $\Lambda$ ($\bar{\Lambda}$) candidate is found, the one with the
largest decay length is retained for further analysis.

In the study of $\psi\to\ssb$ decay, a variable
$\Delta_m=\sqrt{(M_{\Lambda\gamma_{1}}-M_{\Sigma^0})^2+(M_{\bar{\Lambda}\gamma_{2}}-M_{\bar{\Sigma}^0})^2}$
is defined. All possible photon pairs are combined with the selected
$\Lambda$ and $\bar{\Lambda}$ candidates, and the $\gamma_1$ and
$\gamma_2$ candidates, which yield the smallest $\Delta_m$, are taken
as the photons from the $\Sigma^0$ and $\bar{\Sigma}^0$ decays,
respectively.

To suppress backgrounds, the $\Lambda\bar{\Lambda}$ invariant mass,
$M_{\llb}$, is required to be within $[3.05, 3.15]$, $[2.82, 3.02]$,
$[3.63, 3.75]$ and $[3.34, 3.61]$ GeV/$c^2$ for the $J/\psi\to\llb$,
$J/\psi\to\ssb$, $\psi(3686)\to\llb$ and $\psi(3686)\to\ssb$ decays,
respectively. Here the mass window requirements for the individual
decay modes are determined by MC studies. In the decays
$\psi\to\llb$, the $\bar{\Lambda}$ candidate is required to have mass
satisfying
$|M_{\bar{p}\pi^+}-M_{\bar{\Lambda}}|<3\sigma_{M_{\bar{\Lambda}}}$,
where $M_{\bar{\Lambda}}$ is the $\bar{\Lambda}$ nominal mass, and
$\sigma_{M_{\bar{\Lambda}}}$ is the corresponding mass resolution,
which is 2.3 MeV/$c^2$ (4.0 MeV/$c^2$) for the $J/\psi$ ($\psi(3686)$)
decay.  In the decays $\psi\to\ssb$, the $\bar{\Sigma}^0$ candidate is
required to have mass satisfying
$|M_{\bar{p}\pi^+\gamma}-M_{\bar{\Sigma}^{0}}|<3\sigma_{M_{\bar{\Sigma}^{0}}}$,
where $M_{\bar{\Sigma}^0}$ is the $\bar{\Sigma}^0$ nominal mass,
$\sigma_{M_{\bar{\Sigma}^{0}}}$ is the corresponding mass resolution,
which is 4.3 MeV/$c^2$ (6.0 MeV/$c^2$) for the $J/\psi$ ($\psi(3686)$).
The candidates are further required to satisfy
$\theta_{\ssb}>$178$^\circ$ and $\theta_{\ssb}>$178.5$^\circ$ for the
$J/\psi$ and $\psi(3686)$ decays, respectively, where $\theta_{\ssb}$
is the opening angle between the reconstructed $\Sigma^{0}$ and
$\bar{\Sigma}^0$ candidates in the c.m. system.

\section{\bf BACKGROUND ESTIMATION } \label{Sec:bgesti}

To study the backgrounds, the same selection criteria are applied to
the generic inclusive $\psi$ MC samples.
For the decay $J/\psi\to\llb$, the dominant backgrounds are found to
be $J/\psi\to\Lambda\bar{\Sigma}^0+c.c.$, $J/\psi\to\gamma K_sK_s$,
and $J/\psi\to\gamma\eta_{c}$ with the subsequent decay
$\eta_{c}\to\llb$.
For the decay $J/\psi\to\ssb$, the main backgrounds are from
$J/\psi\to\Lambda\bar{\Sigma}^0+c.c.$, $J/\psi\to\gamma\eta_{c}$ with
the subsequent decay $\eta_{c}\to\llb,~\ssb,~\Lambda\bar{\Sigma}^0+c.c.$, 
and $J/\psi\to\Sigma^{0}\bar{\Sigma}^{*0}+c.c.$.
For $\psi(3686)\to\llb$, the potential backgrounds are
$\psi(3686)\to\pi^+\pi^- J/\psi,J/\psi\to p\bar{p}$,
$\psi(3686)\to\ssb$, and $\psi(3686)\to\Lambda\bar{\Sigma}^0 + c.c.$.
For $\psi(3686)\to\ssb$, the dominant backgrounds are from
$\psi(3686)\to\gamma\chi_{cJ}, \chi_{cJ}\to\llb~(J=0,1,2)$ and
$\psi(3686)\to\Xi^{0}\bar{\Xi}^{0},\Xi^{0}\to\Lambda\pi^{0},\bar{\Xi}^{0}\to\bar{\Lambda}\pi^{0}$.
All above backgrounds can be classified into two categories, $i.e.$,
backgrounds with or without $\llb$ in the final state. The former
category backgrounds are expected to produce a peak around the
$\Lambda$/$\Sigma^0$ signal region in the $p\pi^-$/$p\pi^-\gamma$
invariant mass distributions and can be estimated, with the exclusive
MC simulation samples using the decay branching fractions set according
to the PDG~\cite{Ref:pdg}. The additional undetermined decays of 
$\eta_{c}\to\ssb, ~\Lambda\bar{\Sigma}^0+c.c.$  and $\psi(3686)\to\Lambda\bar{\Sigma}^0+c.c.$ 
are estimated using the results from previous experiments for 
charmonium decaying to $B\bar{B}$ states (reference decays)
~\cite{Ref:bes_2006, Ref:bes_2007, Ref:bes_2012},
to be 1 and 0.1 times that for the decay $\eta_{c}\to\llb$ and 0.1 times 
that for $\psi(3686)\to\llb$, respectively. The contributions of other decays
to the peaking background are negligible. The latter category of backgrounds
are expected to be distributed smoothly in the corresponding mass distributions.

The backgrounds from continuum QED processes, $i.e.$ $e^+e^-\to
B\bar{B}$ decays, are estimated with the data samples taken at the
c.m.~energies of 3.08 GeV and 3.65 GeV, which have integrated
luminosities of 30~pb$^{-1}$ and 44~pb$^{-1}$~\cite{Ref:ntot_jpsi,
  Ref:ntot_psip}, respectively.  By applying the same selection
criteria, no event survives in the selection of $J/\psi\to
B\bar{B}$, while in the selection of $\psi(3686)\to B\bar{B}$, only
a few events survive, and no obvious peak is observed in the
$\Lambda$/$\Sigma^{0}$ mass region. The contamination from the QCD
continuum processes can be treated as non-peaking background when
determining the signal yields.

\section{\bf Results} \label{Sec:angdis} 

%
%
\subsection{\bf Branching fractions} \label{Subsec:br} 
\begin{figure}[htbp]
\begin{center}
\includegraphics[height=0.20\textwidth, width=0.235\textwidth]{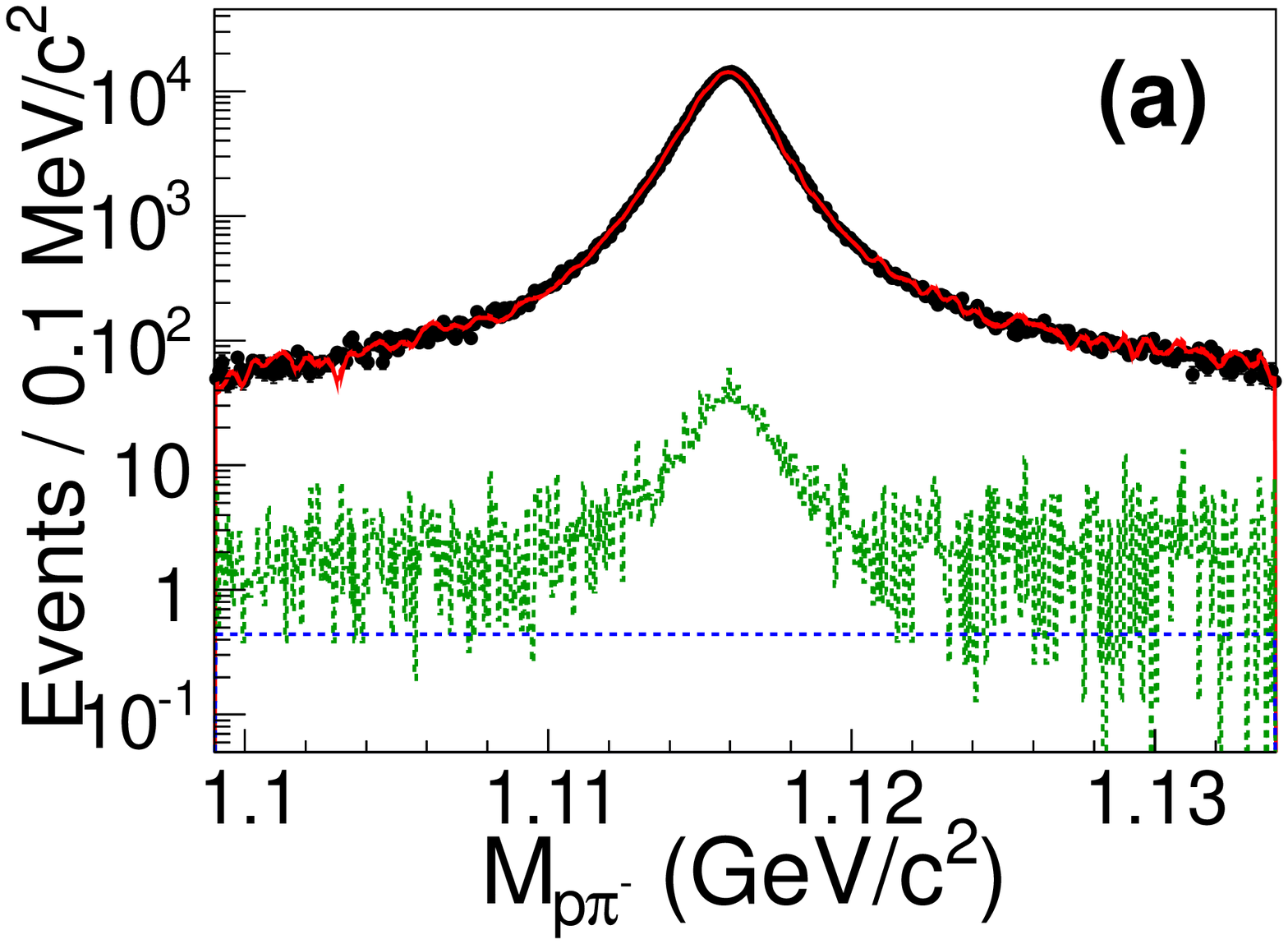}
\includegraphics[height=0.20\textwidth, width=0.235\textwidth]{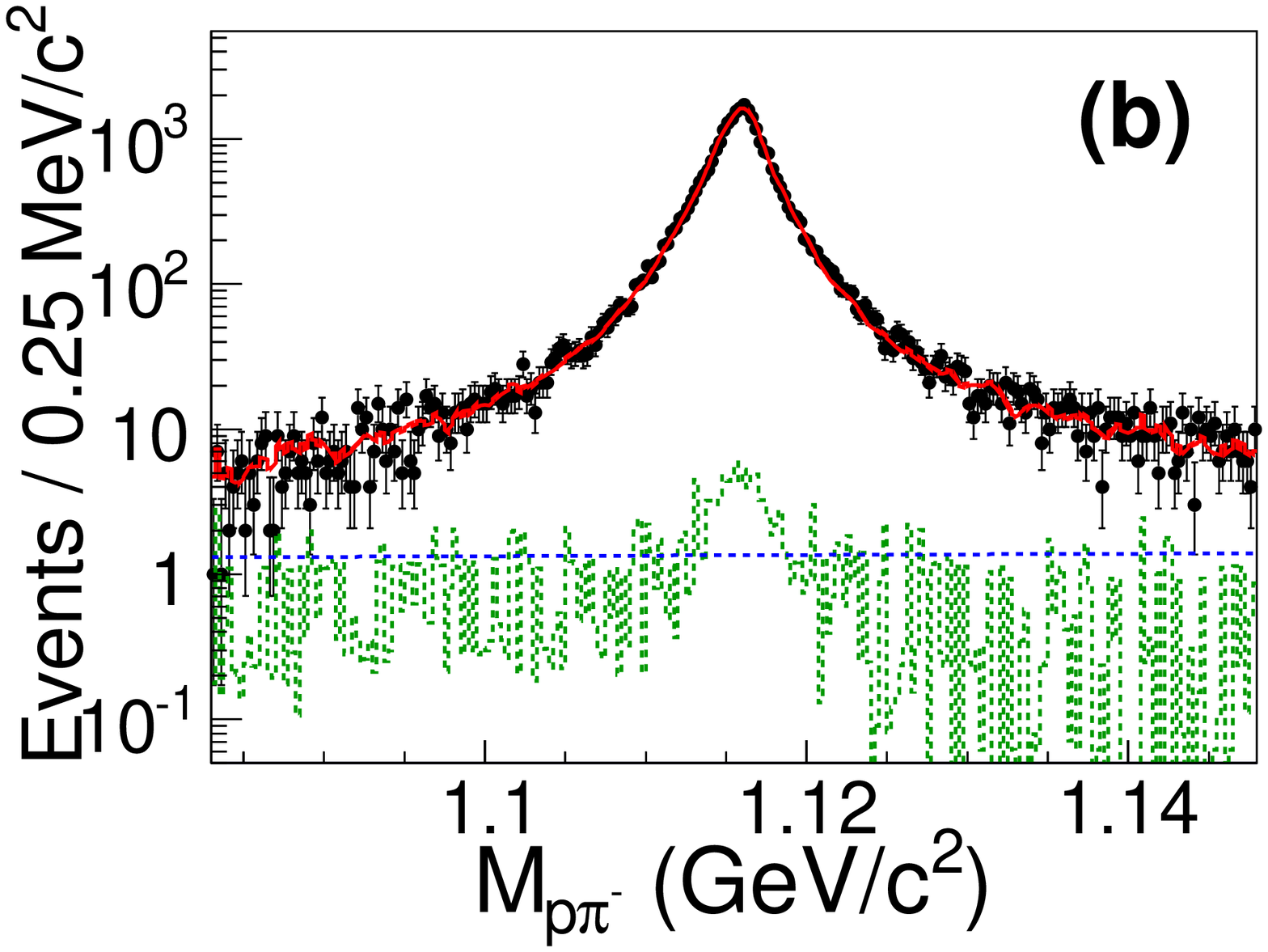}
\includegraphics[height=0.20\textwidth, width=0.235\textwidth]{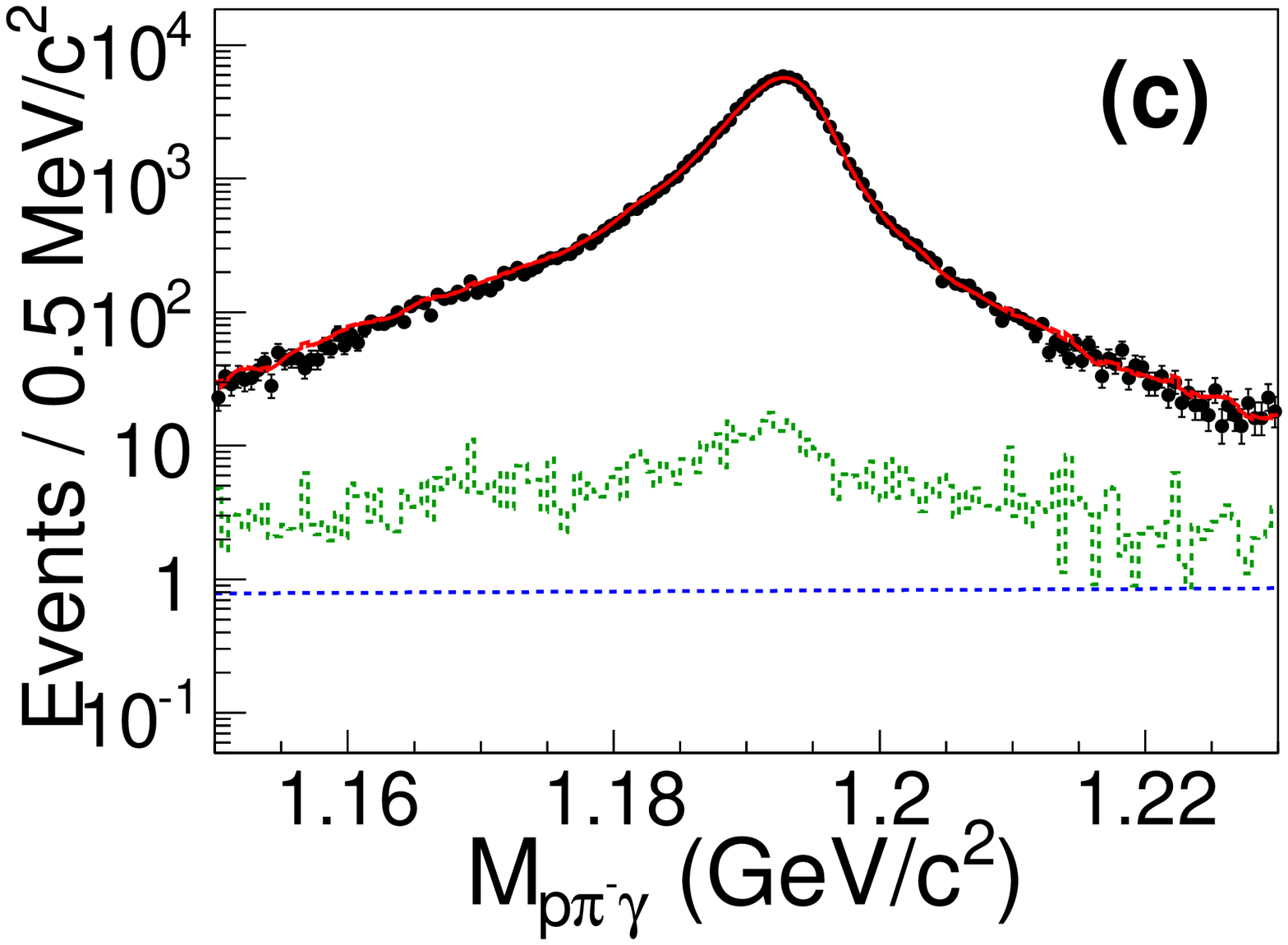}
\includegraphics[height=0.20\textwidth, width=0.235\textwidth]{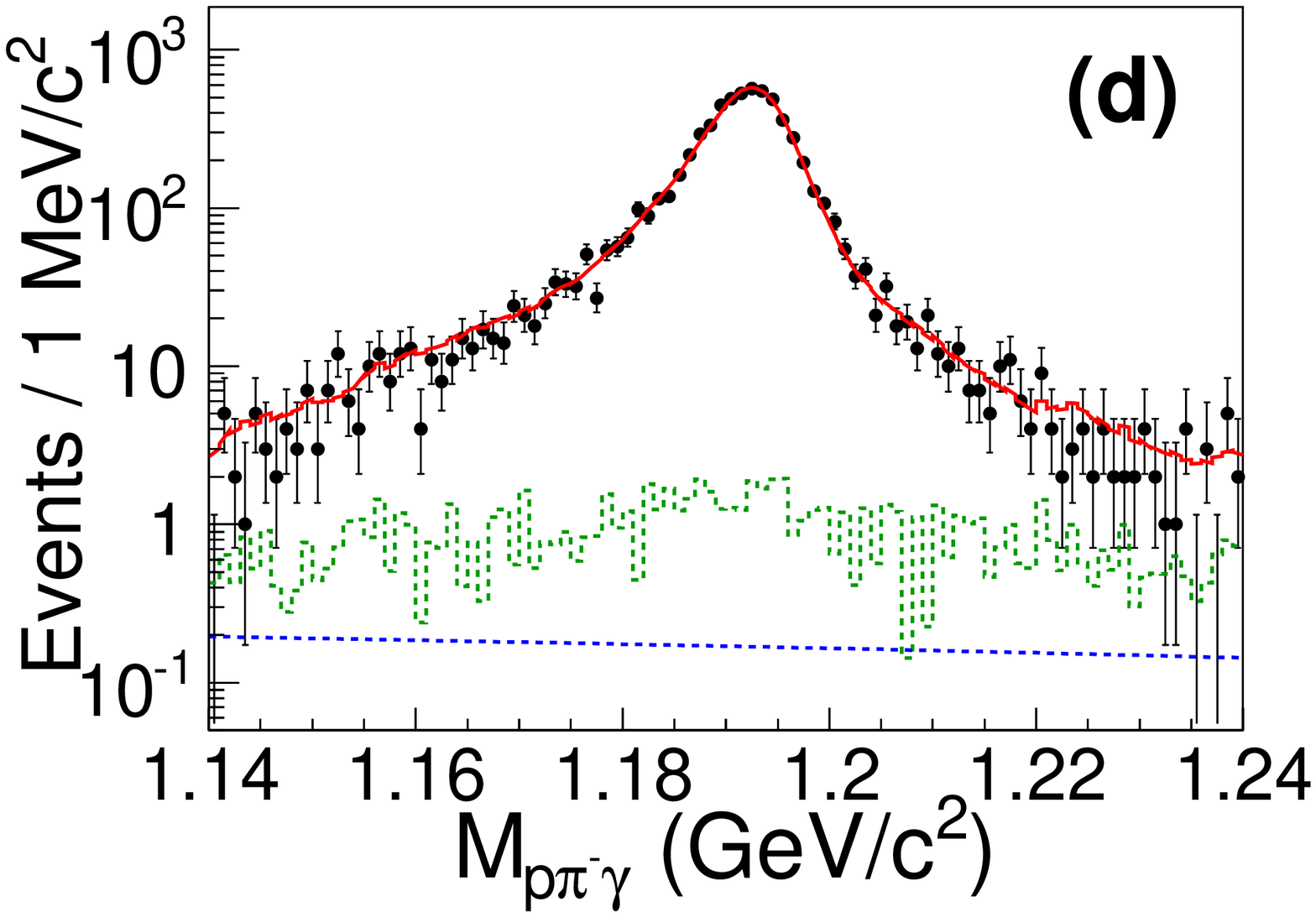}
\caption{(color online) The $M_{p\pi^-}$ distributions for the decays
  (a) $J/\psi\to\llb$ and (b) $\psi(3686)\to\llb$, and the
  $M_{p\pi^-\gamma}$ distributions for the decays (c) $J/\psi\to\ssb$
  and (d) $\psi(3686)\to\ssb$, where the dots with error bars are
  data, the red solid curves are the overall fit results, the green
  dashed histograms are the backgrounds estimated with the exclusive
  MC simulated samples, and the blue dotted line describes the
  remaining backgrounds.}
\label{Fig:mFit}
\end{center}
\end{figure}

With the above selection criteria, the distributions of
$M_{p\pi^-}$/$M_{p\pi^-\gamma}$ in a range of $\pm$8 times the mass
resolution around the $\Lambda$/$\Sigma^{0}$ nominal mass in the
$J/\psi$ and $\psi(3686)$ decays are shown in
Fig.~\ref{Fig:mFit}. Clear $\Lambda$/$\Sigma^{0}$ peaks are observed
with low background. To determine the signal yields, unbinned maximum 
likelihood fits are applied to $M_{p\pi^-}$/$M_{p\pi^-\gamma}$ with the mass 
of $\bar{p}\pi^+$/$\bar{p}\pi^+\gamma'$ restricted to $\pm 3$ times of 
resolution of $\bar{\Lambda}$/$\bar{\Sigma}^0$ nominal mass. 
In the fit, the $\Lambda$/$\Sigma^{0}$
signal shape is described by the simulated MC shape convolved with a
Gaussian function to account for the difference in mass resolution between
data and MC simulation. The peaking backgrounds are described with the
shapes from exclusive MC simulations with fixed magnitudes
according to the branching fractions of background listed in the
PDG~\cite{Ref:pdg}, and the non-peaking backgrounds are described with
second-order polynomial functions with free parameters in the
fit. The fit results are illustrated in Fig.~\ref{Fig:mFit}, and the
corresponding signal yields are summarized in
Table~\ref{Tab:angdis_tab01}.

\begin{table*}[!htb]
  \caption{\label{Tab:angdis_tab01}\small The numbers of observed signal events $N_\text{obs}$, the corrected detection efficiency $\epsilon$, the numbers of peaking backgrounds $N_\text{pk}$, the numbers of smooth backgrounds $N_\text{sm}$, the resultant $\alpha$ values for the angular distributions and the branching fractions $\mathcal{B}$, where the errors are statistical only.}
\begin{tabular}{lcccccc}\hline\hline
Channel        & \multicolumn{1}{c}{$N_\text{obs}$} & \multicolumn{1}{c}{$\epsilon$ (\%)} & \multicolumn{1}{c}{$N_\text{pk}$} & \multicolumn{1}{c}{$N_\text{sm}$} & \multicolumn{1}{c}{$\alpha$}   & \multicolumn{1}{c}{$\mathcal{B}$ ($\times10^{-4}$)}\\
\hline
$J/\psi\to\llb$      & ~$440,675\pm670$ & ~$42.37\pm0.14$ & 1,819 & $154\pm166$ &  ~$\aip$   & ~$\bip$\\
$J/\psi\to\ssb$      & ~$111,026\pm335$ & ~$17.83\pm0.06$ & 820 & $131\pm12$ & ~$\aiip$  & ~$\biip$\\
$\psi(3686)\to\llb$  & ~$31,119\pm187$  & ~$42.83\pm0.34$ & 252 & $352\pm65$ & ~$\aiiip$ & ~$\biiip$\\
$\psi(3686)\to\ssb$  & ~$6,612\pm82$    & ~$14.79\pm0.12$ & 89 & $17\pm5$ & ~$\aivp$  & ~$\bivp$\\
\hline\hline
\end{tabular}
\end{table*}

The branching fractions are calculated using
 \begin{equation}
   \label{equ:br}
   \mathcal{B}(\psi\to B\bar{B})=\frac{N_\text{obs}}{N_{\psi} \cdot \epsilon\cdot\mathcal{B}_{i}},
 \end{equation}
 where $N_\text{obs}$ is the number of signal events minus peaking background;
 $\epsilon$ is the detection efficiency, which is estimated with MC
 simulation incorporating the $\cos\theta$ distributions obtained in
 this analysis and the scale factors to account for the difference in
 efficiency between data and MC simulation as described below;
 $\mathcal{B}_{i}$ is the product of branching fractions for the
 intermediate states in the cascade decay from the PDG~\cite{Ref:pdg};
 and $N_{\psi}$ is the total number of $\psi$ events estimated by
 counting the inclusive hadronic events~\cite{Ref:ntot_jpsi,
   Ref:ntot_psip}. The corresponding detection efficiencies and the
 resultant branching fractions are also summarized in
 Table~\ref{Tab:angdis_tab01}.

%
%
\subsection{\bf Angular distributions} \label{Subsec:angdis}

\begin{figure}[!htb]
\begin{center}
\includegraphics[height=0.20\textwidth, width=0.235\textwidth]{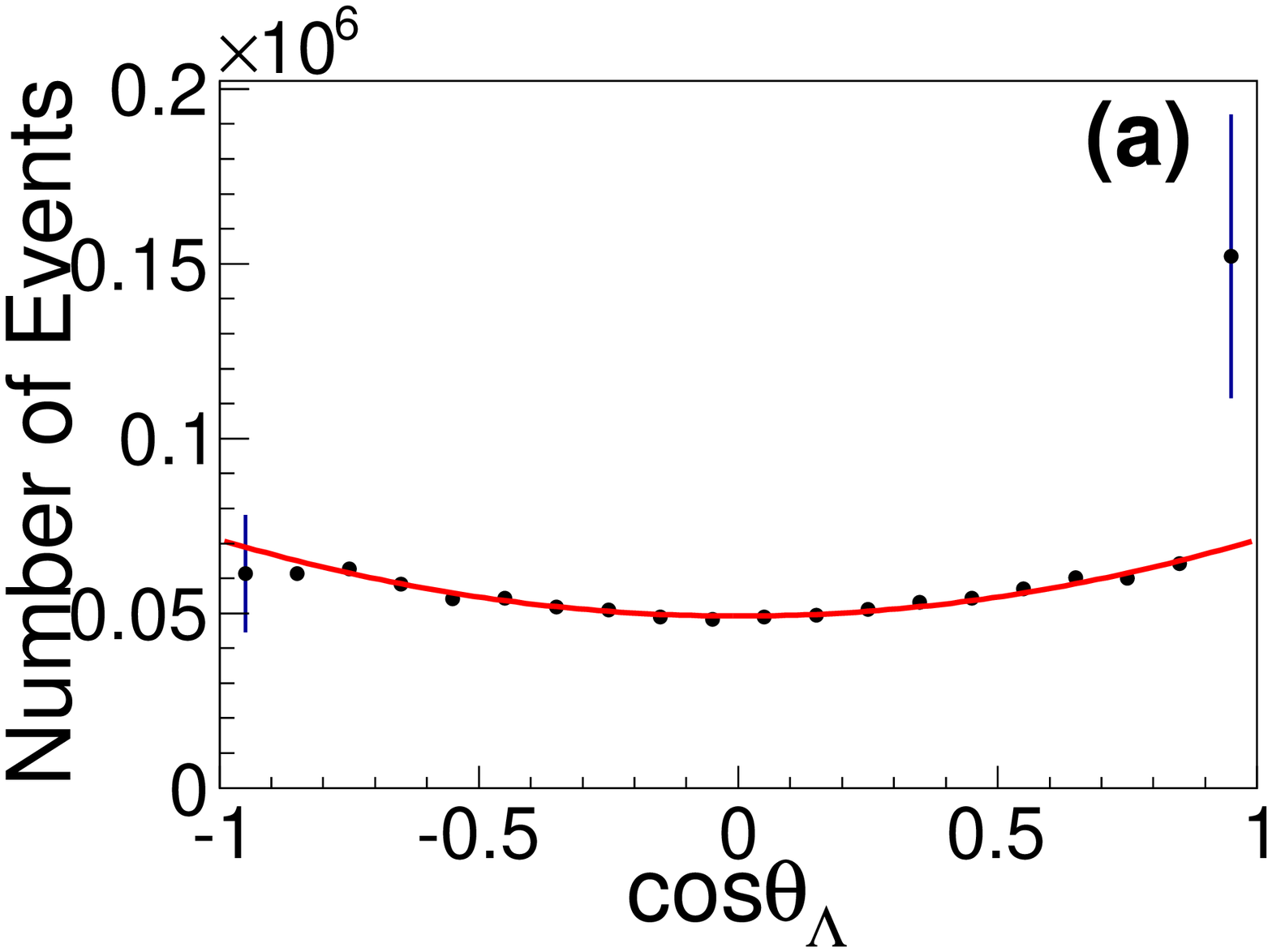}
\includegraphics[height=0.20\textwidth, width=0.235\textwidth]{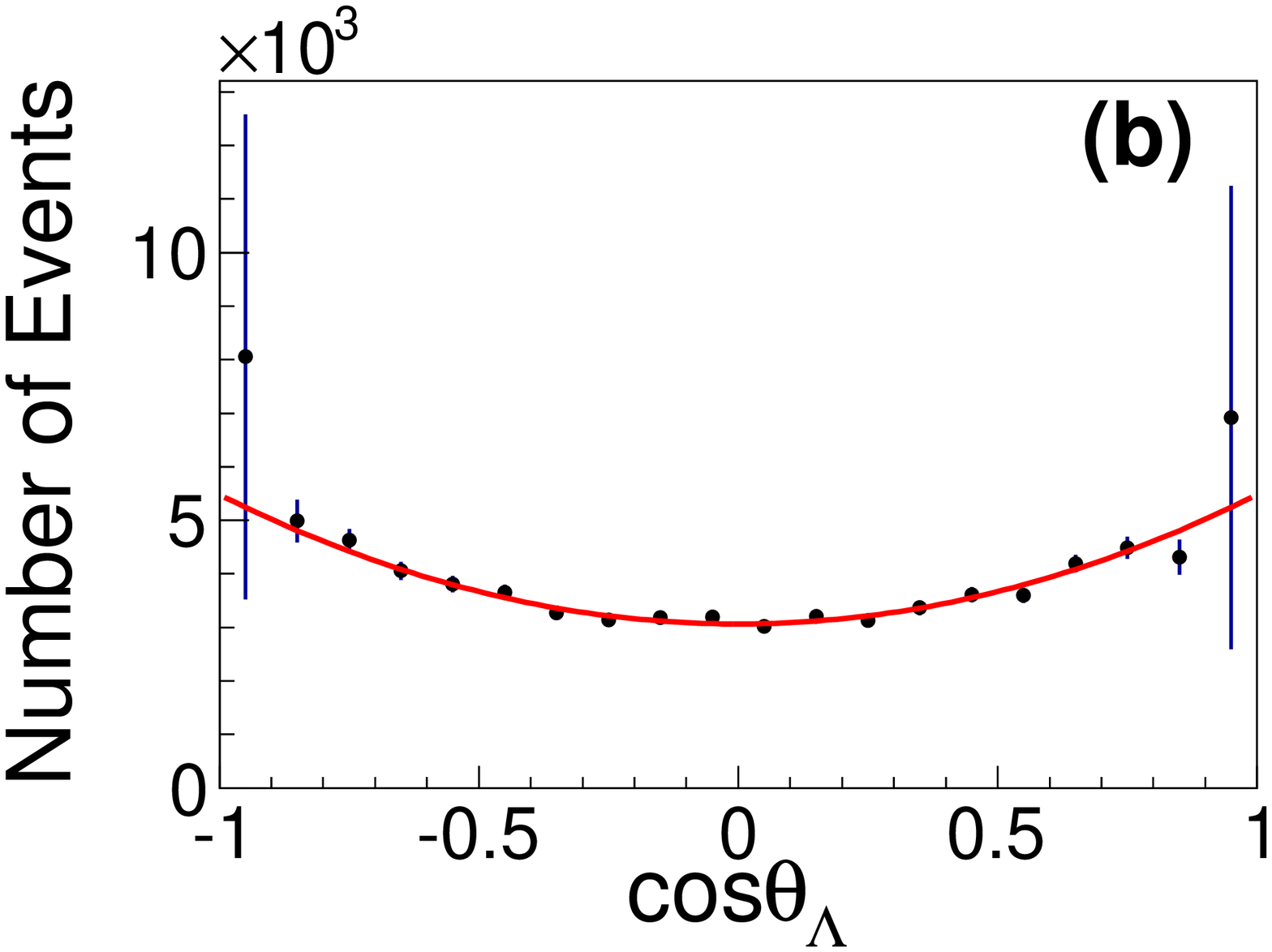}
\includegraphics[height=0.20\textwidth, width=0.235\textwidth]{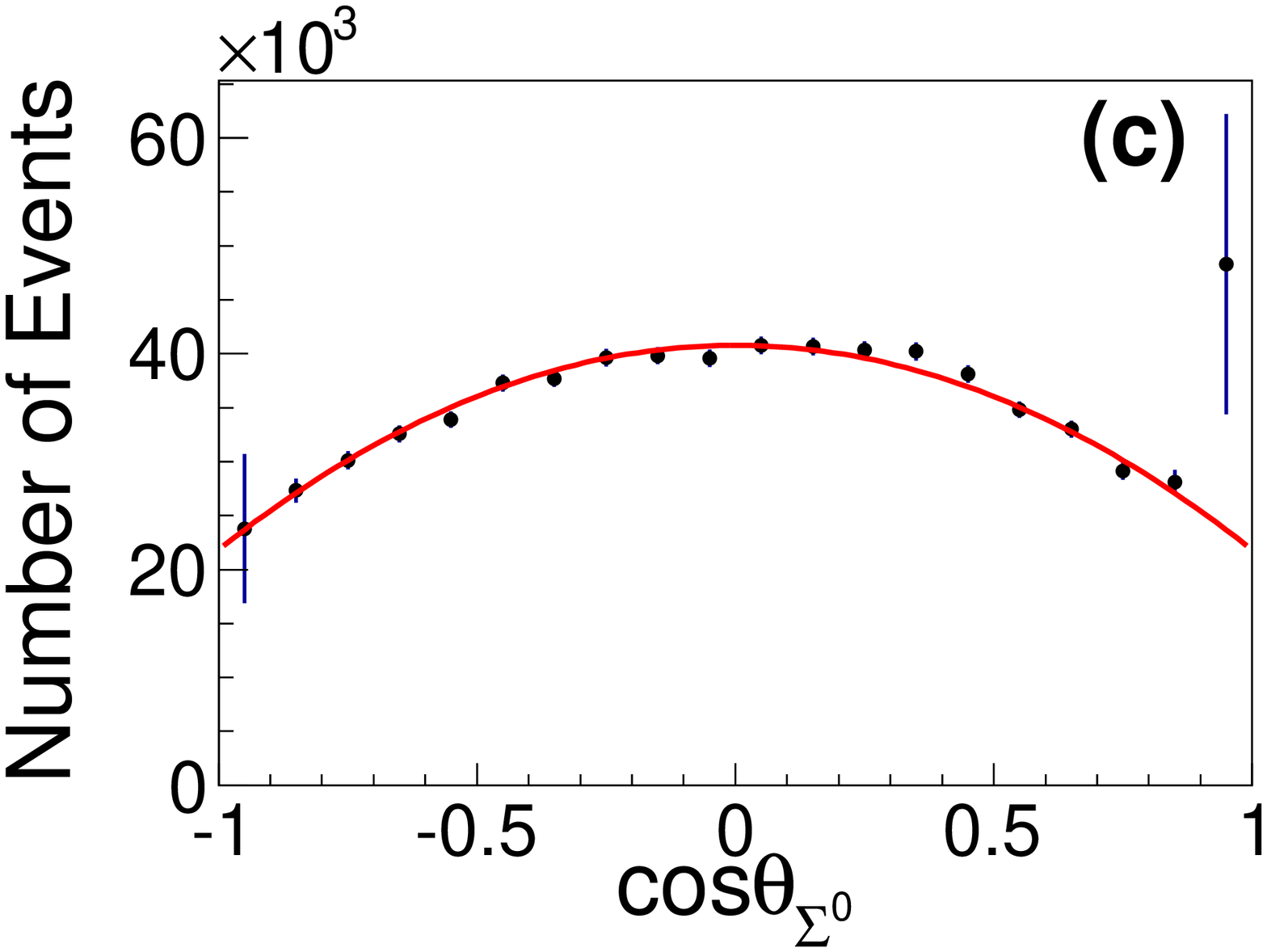}
\includegraphics[height=0.20\textwidth, width=0.235\textwidth]{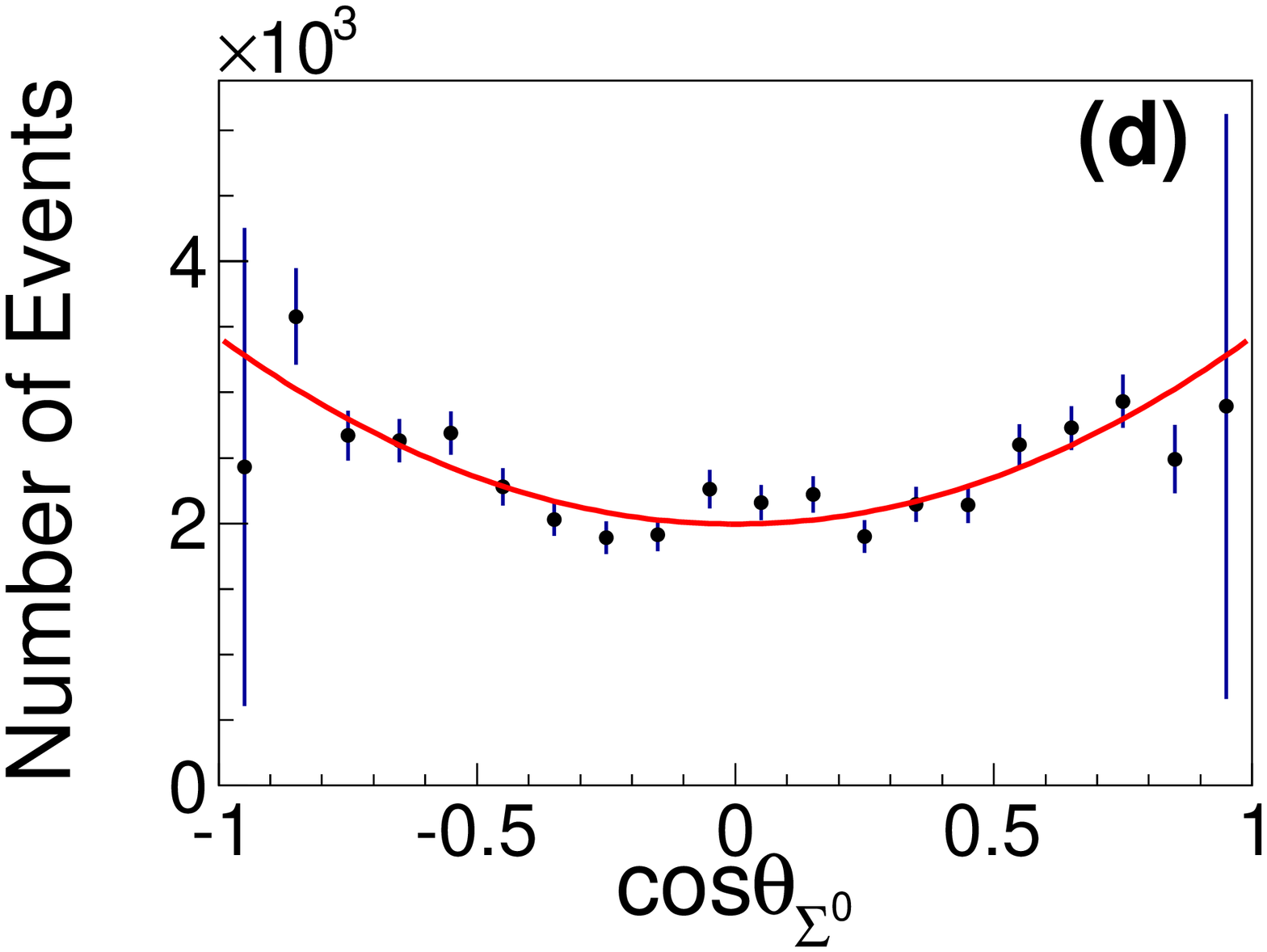}
\caption{(color online) The distributions of efficiency corrected
  polar angle of the baryon for the decays (a) $J/\psi\to\llb$, (b)
  $\psi(3686)\to\llb$, (c) $J/\psi\to\ssb$, and (d)
  $\psi(3686)\to\ssb$, where the dots with error bars are data, and the
  red solid curves are the fit results. }\label{Fig:angularFit}
\end{center}
\end{figure}

The baryon $\cos\theta$ distributions in the c.m.~system corrected by
detection efficiency are shown in Fig.~\ref{Fig:angularFit}, and the
signal yields in each of the 20 bins are determined with the same method as that
in the branching fraction measurements. The detection efficiencies in
each bin are estimated with the signal MC samples and scaled with
correction factors to compensate for the efficiency difference between
data and MC simulation.
The efficiency corrected $\cos\theta$ distributions are fitted with
Eq.~\ref{equ:angdis} with a least squares method, the corresponding fit
results are shown in Fig.~\ref{Fig:angularFit}, and the resultant
$\alpha$ values are summarized in Table~\ref{Tab:angdis_tab01}.

The correction factors used to correct for the efficiency differences
between data and MC simulation as a function of $\cos \theta$ are
determined by studying various control samples, where $\theta$ is 
the polar angle of the hyperon.  The efficiency
differences are due to differences in the efficiencies of charged
particle tracking, photon detection, and hyperon reconstruction.  For
example, the efficiencies related with charged particle tracking and
$\Lambda$ reconstruction are studied with a special control sample of
$\psi\to\llb$ events, where a $\bar{\Lambda}$ tag has been
reconstructed.  Events with two or more charged tracks, in which a $\bar{p}$ 
and $\pi^+$ have been identified using particle
identification, are selected.  The $\bar{\Lambda}$ tag candidate must
satisfy a secondary vertex fit, have a decay length greater than
0.2~cm, and satisfy mass and momentum requirements.  The numbers of
tagged $\Lambda$ events, $N_{tag}$, are obtained by fitting the
$\Lambda$ peak in the distribution of invariant mass recoiling against
the $\bar{\Lambda}$ tag. The numbers of $\Lambda$ signal events,
$N_{sig}$, are obtained by fitting the recoil mass distribution for
events where, in addition, a $\Lambda$ signal is reconstructed on the
recoil side, which requires two oppositely charged tracks that satisfy
a vertex fit and have a decay length greater than 0.2 cm.  The
combined efficiency of charged tracking (proton and pion) and
$\Lambda$ reconstruction is then $N_{sig}/N_{tag}$.  The ratios of the
data and MC simulation efficiencies as a function of $\cos \theta$ are
taken as the correction factors.  The $\bar{\Lambda}$ correction
factors are determined in an analogous way using $\psi\to\llb$ events
with a $\Lambda$ tag.  The overall correction factor in the different
$\cos \theta$ bins is the product of the $\Lambda$ and $\bar{\Lambda}$
correction factors.

In an analogous way, the combined efficiency of photon detection and
$\Sigma^0$ reconstruction is studied with a control sample of
$\psi\to\ssb$ events, which have a $\bar{\Sigma}^0$ tag and an
additional $\Lambda$. Events are selected that have a $\Lambda$ and
$\bar{\Lambda}$ using the same criteria as above and at least one
additional photon.  The $\bar{\Lambda}$ and photon must have an
invariant mass consistent with that of a $\bar{\Sigma}^0$.  The numbers
of tagged $\Sigma^{0}$ events are obtained by fitting the $\Sigma^0$
peak in the distribution of mass recoiling against the
$\bar{\Sigma}^0$ tag.  We then search for another photon and
reconstruct the $\Sigma^0$ by requiring the invariant mass of the
photon and tagged $\Lambda$ be consistent with the $\Sigma^0$ mass.  The
number of events with a $\Sigma^0$ signal divided by the number of
tagged $\Sigma^{0}$ events is the combined efficiency of photon
detection and $\Sigma^0$ reconstruction. The ratios of detection
efficiencies in the different $\cos\theta$ bins between data and MC
simulation, determine the correction factors.  The overall correction
factor in the different $\cos \theta$ bins is the product of the
$\Sigma^0$, $\bar{\Sigma}^0$, $\Lambda$, and $\bar{\Lambda}$
correction factors.

%
%
\section{\bf SYSTEMATIC UNCERTAINTY} \label{Sec:sys}

\subsection{Branching Fraction}\label{Subsec:brsys}
Systematic uncertainties in the branching fraction measurements are
mainly due to the differences of detection efficiency and resolution
between data and MC simulation.  The sources of uncertainty related
with the detection efficiency include charged tracking, photon
detection, and $\Lambda$/$\Sigma^0$ reconstruction.  The sources of
uncertainty due to the resolution difference include the $M_{\llb}$
and $M_{\bar{\Lambda}}/M_{\bar{\Sigma}^{0}}$ mass requirements, and the
opening angle $\theta_{\ssb}$ requirement in the decays $\psi\to\ssb$.
Additional uncertainty sources including the model of the baryon polar
angular distribution, the fit procedure, the decay branching fractions
of $\Lambda$/$\Sigma^0$ states and the total number of $\psi$ events
are also considered. All of systematic uncertainties are studied in
detail as discussed in the following:

\begin{enumerate}

\item As described above, the detection efficiencies related with the
  tracking, photon detection, and $\Lambda$/$\Sigma^0$ reconstruction
  are corrected bin-by-bin in $\cos\theta$ to decrease the difference
  between data and MC simulation. The overall correction factors,
  which are determined with control samples are
  $0.9974\pm0.0041$, $0.9936\pm0.0064$, $0.980\pm0.011$, and
  $0.954\pm0.022$ for the decays $J/\psi\to\llb$, $J/\psi\to\ssb$,
  $\psi(3686)\to\llb$ and $\psi(3686)\to\ssb$, respectively. To
  estimate the corresponding uncertainties, the correction factors are
  changed by $\pm$1 standard deviations, and the
  resultant changes in the branching fractions are taken
  as the systematic uncertainties.

\item The uncertainties related with the $M_{\llb}$ requirement are
  estimated by varying the mass requirement edges by $\pm10$
  MeV/$c^2$. The uncertainties related with the
  $\bar{\Lambda}/\bar{\Sigma}^{0}$ mass requirement are estimated by
  changing the requirement by $\pm 1$ times the mass resolution. The
  uncertainties due to the requirement on the opening angle
  $\theta_{\ssb}$ in the decays $\psi\to\ssb$ are estimated by
  changing the requirement to be 175$^\circ$. The relative changes in
  the branching fractions are individually taken as the
  systematic uncertainties.

\item MC simulations indicate that the detection efficiencies depend
  on the distributions of baryon polar angular $\cos\theta$. In the
  analysis, the measured $\alpha$ values are used for the
  $\cos\theta$ distributions in the MC simulation. Alternative MC
  samples are generated by changing the $\alpha$ values by $\pm1$
  standard deviations and are used to estimate the detection
  efficiencies. The resultant changes in the detection efficiencies
  with respect to their nominal values are taken as the systematic
  uncertainties.

\item The sources of systematic uncertainty associated with the fit
  procedure include the fit range, the signal shape and the modeling
  of backgrounds.  The uncertainties related with the fit range are
  estimated by changing the range by $\pm1$ times the mass resolution
  for the fits.  The signal shapes are modeled with the signal MC
  simulated shapes convolved with a Gaussian function in the nominal
  fit. The corresponding uncertainties are estimated with alternative
  fits with different signal shapes, $i.e.$, a Breit-Wigner function
  convolved with a Gaussian function for $\Lambda$ and with a Crystal
  Ball function~\cite{Ref:cball} for $\Sigma^{0}$, where the Gaussian
  function and Crystal Ball function represent the corresponding
  mass resolutions.  The uncertainties related with the peaking
  backgrounds, which are estimated with the exclusive MC samples in
  the nominal fits, are studied by changing the branching fractions of
  the individual background, or by changing the branching fractions for 
  the reference decays which the estimated branching fractions for the 
  undetermined backgrounds are based on,
  by $\pm1$ times their uncertainties from the PDG~\cite{Ref:pdg}. 
  The uncertainties associated with the
  non-peaking backgrounds are estimated with alternative fits by
  replacing the second order polynomial function with a first order
  polynomial function.  The resultant changes from the above changes
  in the signal yields are taken individually as the systematic
  uncertainties.

\item The uncertainties related with the branching fractions of baryon
  and anti-baryon decays are taken from the PDG~\cite{Ref:pdg}. The total
  numbers of $\psi$ events are obtained by studying the inclusive
  hadronic events, and their uncertainties are 0.6\% and 0.7\% for the
  $J/\psi$ and $\psi(3686)$ data samples~\cite{Ref:ntot_jpsi,
    Ref:ntot_psip}, respectively.

\end{enumerate}

The various systematic uncertainties in the branching fraction
measurements are summarized in Table~\ref{Tab:sys_tab02}. The total
systematic uncertainties are obtained by summing the individual values
in quadrature.

\begin{table}[!htb]
\caption{\label{Tab:sys_tab02}\small Systematic uncertainties in the measurement of branching fractions (\%).}
\begin{center}
\begin{tabular}{lcccc}\hline\hline
                                          & \multicolumn{2}{c}{$J/\psi$}  &  \multicolumn{2}{c}{$\psi(3686)$}  \\ \cline{2-3} \cline{4-5}
                                          & ~~$\llb$~~ & ~~$\ssb$~~ & ~~$\llb$~~ & ~~$\ssb$~~ \\\hline
    Efficiency correction                 & 0.5  & 0.7          & 1.2 & 2.3\\
    $M_{\llb}$ requirement                & 0.1 & 0.1           & 0.1 & 0.2\\
    $\bar{\Lambda}/\bar{\Sigma}^{0}$ mass requirement      & 0.1 & 0.3           & 0.3 & 0.2\\
    $\theta_{\ssb}$ requirement           &  $-$  & 0.3 &  $-$  & 0.2\\
    Baryon polar angle     & 0.8 & 0.9           & 2.0 & 3.1\\
    Fit range              & 0.1 & 0.1           &  0.2 & 0.2 \\
    Signal shape                 & 0.1 & 0.3           &  0.1 & 0.2 \\
    Peaking bkg.                 & 0.3 & 0.4           & 0.3 & 1.2\\
    Non-peaking bkg.                 & 0.1 & 0.1           & 0.3 & 0.2\\
    Branching fractions			          & 1.2 & 1.2           & 1.2 & 1.2\\
    $N_{J/\psi}/N_{\psi(3686)}$           & 0.6 & 0.6           & 0.7 & 0.7 \\\hline
    Total                                 & 1.7 & 1.9           & 2.8 & 4.3 \\
    \hline\hline
    \end{tabular}
\end{center}
\end{table}
%

\subsection{Angular Distribution }\label{Subsec:angdissys}

The sources of systematic uncertainties in the baryon polar angular
measurements include the signal yields in different $\cos\theta$
intervals and the $\cos\theta$ fit procedure. The MC statistics and 
correction errors are already included in the error referred to as ``statistical".
\begin{enumerate}

\item In the polar angular measurements, the signal yield in a given
  $\cos\theta$ interval is obtained with the same fit method as that
  used in the branching fraction measurements. The uncertainties of
  the signal yield in each $\cos\theta$ bin are mainly from the fit
  range, the signal shape and the background modeling.
   We individually estimate the uncertainty of the signal yield in each
  $\cos\theta$ interval with the same methods as those used in the
  branching fraction measurements for the different uncertainty
  sources, and then repeat the $\cos\theta$ fit procedure with the
  changed signal yields. The resultant changes in the $\alpha$ values
  with respect to the nominal values are taken as systematic
  uncertainties.

\item The sources of systematic uncertainty related to the
  $\cos\theta$ fit procedure include the fit range and the number of
  bins in the $\cos\theta$ distribution. We repeat the fit procedures
  with the alternative fit range $[-0.9, 0.9]$ and alternative
  number of bins (40). The resultant changes of $\alpha$ values are
  taken as the systematic uncertainties.

\end{enumerate}

The individual absolute uncertainties in the polar angular
distribution measurements are summarized in Table~\ref{Tab:sys_tab01}.
The total systematic uncertainties are obtained by summing the
individual values in quadrature.

\begin{table}[!htb]
\caption{\label{Tab:sys_tab01}\small Absolute systematic uncertainties in the measurement of $\alpha$.}
\begin{center}
\begin{tabular}{lcccc}\hline\hline
                                & \multicolumn{2}{c}{$J/\psi$}& \multicolumn{2}{c}{$\psi(3686)$} \\ \cline{2-5}
                                &~~$\llb$~~ & ~~$\ssb$~~~ & ~~~$\llb$~~ & ~~$\ssb$~~ \\\hline
    Mass fit range     & 0.001 & 0.001           &  0.003 & 0.005 \\
    Signal shape       & 0.001 & 0.002           &  0.001 & 0.003 \\
    Peaking bkg.       & 0.006 & 0.005 & 0.006 & 0.015 \\
    Non-peaking bkg.   & 0.002 & 0.001 & 0.004 & 0.002 \\
    $\alpha$ fit range & 0.001 & 0.003 & 0.007 & 0.019 \\
    Number of bins      & 0.004 & 0.005 & 0.001 & 0.024 \\ \hline
    Total                      & 0.008 & 0.008 & 0.011 & 0.035 \\
    \hline\hline
    \end{tabular}
\end{center}
\end{table}

%
%
\section{\bf Summary} \label{Sec:sum} 
In summary, using the data
samples of $\Nj$ $J/\psi$ events and $\Np$ $\psi(3686)$ events
collected with the BESIII detector at the BEPCII collider, the
$J/\psi$ and $\psi(3686)$ decaying into $\llb$ and $\ssb$ pairs are
studied. The decay branching fractions and $\alpha$ values are
measured, and the results are summarized in Table~\ref{Tab:sum_tab01}.
The branching fractions for $J/\psi$ decays are in good agreement with
the results of BESII~\cite{Ref:bes_2006} and BaBar~\cite{Ref:babar_2007}
experiments, and those for $\psi(3686)$ decays are in agreement with
the results of CLEO~\cite{Ref:cleo_2005}, BESII~\cite{Ref:bes_2007} and
S.~Dobbs {\it et al.}~\cite{Ref:cleo_2014} with a maximum of 2 times of 
standard deviations. The earlier experimental results~\cite{Ref:markii_1984, Ref:dm2_1987, Ref:bes_1998, Ref:bes_2001}
have significant differences with those of this analysis. The precisions of our branching fraction results are 
much improved than those of previous experiments listed in Table~\ref{Tab:br_list}.
The $\alpha$ values in the decays
$\psi(3686)\to\llb$ and $\psi(3686)\to\ssb$ are measured for the first
time, while those of $J/\psi\to\llb$ and $J/\psi\to\ssb$ decays are of
much improved precision compared to previous measurements.  It is
worth noting that the $\alpha$ value in the decay $J/\psi\to\ssb$ is
negative, which confirms the results in Ref.~\cite{Ref:bes_2006}.
\begin{table}[!htb]
\begin{center}
\caption{\label{Tab:sum_tab01}\small Results for measured $\alpha$ values and branching fractions $\mathcal{B}$ in this analysis. The first uncertainties are statistical, and the second are systematic.}
\footnotesize
\begin{tabular}{lcc}\hline\hline
Channel  &  $\alpha$  & $\mathcal{B}$ ($\times10^{-4}$) \\
\hline
$J/\psi\to\llb$     & \multicolumn{1}{c}{$\ai$}   & \multicolumn{1}{c}{$\bi$}   \\
$J/\psi\to\ssb$     & \multicolumn{1}{c}{$\aii$}  & \multicolumn{1}{c}{$\bii$}  \\
$\psi(3686)\to\llb$ & \multicolumn{1}{c}{$\aiii$} & \multicolumn{1}{c}{$\biii$} \\
$\psi(3686)\to\ssb$ & \multicolumn{1}{c}{$\aiv$}  & \multicolumn{1}{c}{$\biv$}  \\
\hline\hline
\end{tabular}
\end{center}
\end{table}

To test the ``12\% rule'', we also obtain the Q values to be
$\frac{\mathcal{B}(\psi(3686)\to\llb)}{\mathcal{B}(J/\psi\to\llb)}=~
$($\qi$)\% and
$\frac{\mathcal{B}(\psi(3686)\to\ssb)}{\mathcal{B}(J/\psi\to\ssb)}=~
$($\qii$)\%, where the common systematic uncertainties between
$J/\psi$ and $\psi(3686)$ decays are cancelled. The Q values are of
high precision, and differ from the expectation from pQCD by
more than 3 standard deviations.

\section{Acknowledgments}
The BESIII collaboration thanks the staff of BEPCII and the IHEP computing center for their strong support. This work is supported in part by National Key Basic Research Program of China under Contract Nos. 2009CB825200, 2015CB856700; National Natural Science Foundation of China (NSFC) under Contracts Nos. 10905034, 10935007, 11125525, 11235011, 11322544, 11335008, 11425524; the Chinese Academy of Sciences (CAS) Large-Scale Scientific Facility Program; the CAS Center for Excellence in Particle Physics (CCEPP); the Collaborative Innovation Center for Particles and Interactions (CICPI); Joint Large-Scale Scientific Facility Funds of the NSFC and CAS under Contracts Nos. 11179007, U1232106, U1232201, U1332201; Natural Science Foundation of Shandong Province under Contract No. ZR2009AQ002; CAS under Contracts Nos. KJCX2-YW-N29, KJCX2-YW-N45; 100 Talents Program of CAS; National 1000 Talents Program of China; INPAC and Shanghai Key Laboratory for Particle Physics and Cosmology; German Research Foundation DFG under Contract No. Collaborative Research Center CRC-1044, FOR-2359; Istituto Nazionale di Fisica Nucleare, Italy; Joint Funds of the National Science Foundation of China under Contract No. U1232107; Ministry of Development of Turkey under Contract No. DPT2006K-120470; Russian Foundation for Basic Research under Contract No. 14-07-91152; The Swedish Resarch Council; U. S. Department of Energy under Contracts Nos. DE-FG02-04ER41291, DE-FG02-05ER41374, DE-SC0012069, DESC0010118; U.S. National Science Foundation; University of Groningen (RuG) and the Helmholtzzentrum fuer Schwerionenforschung GmbH (GSI), Darmstadt; WCU Program of National Research Foundation of Korea under Contract No. R32-2008-000-10155-0.

\end{document}